\documentclass[aps,pre,twocolumn,showpacs]{revtex4}
\usepackage{epsfig}
\usepackage{times}
\begin{document}

\title{Evolutionary Prisoner's Dilemma game on the Newman-Watts networks}

\author{Jeromos Vukov$^1$, Gy\"orgy Szab\'o$^2$, and Attila Szolnoki$^2$}
\affiliation {$^1$Department of Biological Physics, E\"otv\"os
University, H-1117 Budapest, P\'azm\'any P. stny. 1/A., Hungary  \\
$^2$Research Institute for Technical Physics and Materials Science
P.O. Box 49, H-1525 Budapest, Hungary}

\begin{abstract}
Maintenance of cooperation was studied for a two-strategy evolutionary Prisoner's Dilemma game where the players are located on a one-dimensional chain and their payoff comes from games with the nearest and next-nearest neighbor interactions. The applied host geometry makes possible to study the impacts of two conflicting topological features. The evolutionary rule involves some noise affecting the strategy adoptions between the interacting players. Using Monte Carlo simulations and the extended versions of dynamical mean-field theory we determined the phase diagram 
as a function of noise level and a payoff parameter. 
The peculiar feature of the diagram is changed significantly when the connectivity structure is extended by extra links as suggested by Newman and Watts. 
\end{abstract}

\pacs{89.65.-s, 05.50.+q, 02.50.+Le, 87.23.Ge}

\maketitle

Maintenance of cooperative behavior among selfish individuals in biological and social systems is a progressively studied and challenging problem \cite{axelrod_84}. The celebrated Prisoner's Dilemma (PD) game \cite{gintis_00} is a widely applied mathematical model illustrating the conflict between the individually rational (selfish) and globally useful (cooperative) behavior.

In the original two-person (one-shot) game the players can follow one of the two strategies called cooperation ($C$) and defection ($D$). The player's payoff (or fitness) depends on their choice that is determined by the elements of a payoff matrix. To be more specific, for mutual cooperation each player receives the reward $R$, two defectors receive the punishment $P$, whilst a cooperator and defector receive the suckers payoff $S$ and the temptation $T$ (to choose defection), respectively. For the PD game the payoffs satisfy the ranking: $T > R > P > S$. According to the assumption of the traditional game theory, players make rational decision to maximize their own income. Consequently, they should choose defection independently from the other player's decision. For the iterated PD game an additional constraint (namely $2 R > T + P$) is assumed to provide the highest total income for mutual cooperation. 

In the spatial evolutionary PD games $N$ players are distributed on a lattice with periodic boundary conditions. Following one of the above mentioned pure strategies each player plays a game with her neighbors. These games are repeated meanwhile the players are allowed to modify their strategies in a way defined by the strategy update rule. For example, in the model suggested by Nowak and May \cite{nowak_ijbc93} the players were located on the sites of a square lattice, they played a PD game with all neighbors for discrete times $t=1,2, \ldots$, and before the next time step each player has adopted the strategy of those player who received the highest accumulated payoff in her neighborhood (including herself). In these cellular automaton type models the coexistence of the $C$ and $D$ strategies is provided by a frozen or oscillating pattern due to the dynamical balance between the opposite invasion processes affected by the payoffs. For synchronized strategy updates $C$ invasion can occur along the horizontal or vertical interfaces separating the domains of the $C$ and $D$ strategies. On the contrary, $D$ invasions are favored along the irregular interfaces where defectors can exploit many neighboring cooperators. The average density of cooperators is reduced drastically for asynchronous strategy updates \cite{nowak_ijbc94} as well as for the introduction of additional noise including irrational strategy adoptions \cite{blume_geb03} (for a recent survey see Ref.~\cite{szabo_pr07}).

In the last years the spatial models were generalized by locating players on the sites of a network (graph) whose edges connect each player with her co-players (neighbors) \cite{ebel_pre02,kim_pre02,duran_pd05,santos_prl05,ohtsuki_jtb06a}. Investigations were also extended to different versions of small-world structures created from a one- or two-dimensional lattice by rewiring some portion of connections \cite{abramson_pre01,masuda_pla03,santos_pre05,guan_cpl06,wu_cpl06} or by adding extra links to the spatial structure \cite{kim_pre02,wu_pre05,ren_pre07,fu_epjb07}. The systematic comparison of the cited results is difficult because of the wide variety of dynamical rules used in the mentioned studies. Nevertheless, it is concluded generally that the introduction of inhomogeneities supports the maintenance of cooperation among selfish individuals
\cite{ren_pre07,guan_pre07b}. An important progress has been made by 
Santos {\it et al.}, who described a mechanism that provides relevant advantage for cooperators in the presence of connected hubs (players with a large number of neighbors) \cite{santos_prl05}.

Further exploring the possible effects of the interaction topology on the cooperation level we
study a simple host structure where two conflicting topological features can be detected.
It is well known that cooperators die out in one-dimensional structure ($z=2$) for all stochastic dynamical rules favoring the adoption of the more successful strategies \cite{abramson_pre01,masuda_pla03,santos_prl05}. Similar result is predicted by the mean-field analysis \cite{szabo_pr07} describing a system where the co-players are chosen randomly for both playing games and learning strategies. The former results explain why the Monte Carlo (MC) simulations have indicated the extinction of cooperators for a sufficiently large $z$ dependent on the evolutionary rule(s) and payoffs \cite{santos_prl05,tang_epjb06}. In contrast, the systematic investigations  have elucidated that the cooperation can be maintained for random sequential updates  because the one-site overlapping triangles (in regular connectivity structure at $z=4$) support the spreading of cooperation in the low noise limit \cite{szabo_pre05}. According to the MC simulations (performed on many different two- and three-dimensional lattices and other regular networks) this feature remains valid for the investigated spatial and non-spatial connectivity structures if the overlapping triangles span the whole graph. Following this avenue,
a more rigorous analysis of this feature requires to study those regular structures which cannot be spanned completely by the overlapping triangles or on which the overlapping triangles form a one-dimensional structure favoring an evolution towards homogeneous states via a domain growing process. Now our attention is focused on the latter case by considering a simple connectivity structure combining the mentioned conflicting topological features that become relevant in the low noise limit.

For this purpose we study a two-strategy evolutionary PD game on a one-dimensional lattice where the sites are connected to $z=4$ neighbors. The systematic investigation of the effect of noise and payoffs on the density of cooperators in the stationary states requires accurate knowledge about what happens on a host graph where the overlapping triangles span the whole system but the one-dimensional feature is preserved. Due to this curiosity the resultant phase diagram differs significantly from those studied previously as detailed below. Furthermore we show that the mentioned unique behavior is modified significantly when the small-world effect is switched on by adopting the construction suggested by Newman and Watts \cite{newman_pla99}.

In our model $N$ players are located on the nodes $x$ of a network and they can follow the $C$ or the $D$ strategies. The spatial strategy distribution is described by
a set of two-state site variables, i.e., ${\bf s}_x=C$ or $D$. For
later convenience the local states are denoted by two-dimensional unit vectors
\begin{equation}
\label{eq:cd} C= \left( \matrix{1 \cr 0 \cr }\right) \;\;
\mbox{and}\;\; D= \left( \matrix{0 \cr 1 \cr }\right) \;,
\end{equation}
and in this notation the payoffs can be expressed with the aid of
simple matrix algebra. The normalized income of the player at the
site $x$ obeys the following form:
\begin{equation}
\label{eq:tpo} U_x=\frac{4}{\left| \Omega_x \right|}\sum_{y \in
\Omega_x } {\bf s}^{+}_x {\bf A} \cdot {\bf s}_y\,\,,
\end{equation}
where ${\bf s}^{+}_x$ is the transpose of the state vector ${\bf s}_x$, and the summation runs over all the neighbors ($\Omega_x$)
of player $x$ ($\left| \Omega_x \right|$ refers to the number of
neighbors surrounding the site $x$). The pre-factor in
Eq.~\ref{eq:tpo} is applied to help the comparison of the
resultant phase diagrams with other systems analyzed in previous
studies~\cite{perc_epl06,guan_epl06,szabo_pre05}. Due to its
simplicity we use the rescaled payoff matrix suggested by Nowak
{\it et al.} \cite{nowak_ijbc93}, i.e.:
\begin{equation}
{\bf A}=\left( \matrix{1 & 0 \cr
                       b & 0 \cr} \right)\;, \;\; 1 < b < 2\,.
\end{equation}
In this notation $1<b<2$ denotes the strength of temptation to
exploit the neighboring cooperator(s).

The time evolution of the strategy distribution is modelled by the
so-called pairwise comparison dynamics when a randomly chosen
player $x$ adopts the strategy of her randomly chosen
neighbors ($y$) with a probability depending on the payoff
difference as
\begin{equation}
\label{eq:update} W[{\bf s}_x \leftarrow {\bf s}_y] = {1 \over 1 +
 \exp {[(U_x-U_y)/K]} } \;,
\end{equation}
where $K$ characterizes the strength of noise in the strategy
adoption process allowing the players to make irrational decisions
too \cite{szabo_pre98,perc_njp06a,traulsen_jtb07}.
In other words $x$ player can adopt the strategy of $y$ player even if 
$U_x > U_y$.
Starting from a random initial state the
above strategy adoption process is repeated until the system
reaches a stationary states characterized by the average density
$\rho$ of cooperators dependent on the noise $K$, temptation $b$,
and connectivity structure. Three different stationary states can
be distinguished when the number of sites tends to infinity.
In the first case cooperators dominate the whole system ($\rho=1$) if $b < b_{c1}$.
On the contrary, for sufficiently high temptation ($b> b_{c2}$)
only the defectors can survive. In general, there exists a
coexistence region ($0<\rho<1$ for $b_{c1}< b < b_{c2}$) where
both C and D survive in the stationary state. In order to quantify
both critical values of $b$ our analysis is extended to the region
of $b<1$ belonging to the so-called Stag Hunt games representing
another social dilemma \cite{skyrms_03}.

\begin{figure}[h]
\centerline{\epsfig{file=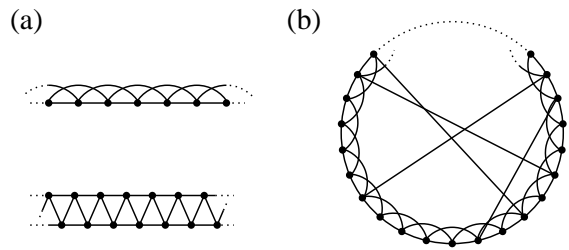,width=8cm}}
\caption{\label{fig:struct} Different connectivity structures for
which the evolutionary PD game is studied. (a) One-dimensional
chain with nearest and next-nearest neighbor interactions ($z=4$).
The second (equivalent) graph below illustrates the overlapping
triangle structure. (b) One-dimensional chain with Newman-Watts
small-world modification.}
\end{figure}

Now our analysis is restricted to two types of connectivity
structures illustrated in Fig.~\ref{fig:struct}. The first
structure is a linear chain for nearest and next-nearest
interactions with periodic boundary conditions. The second
structure is created from the first one by adding new links to the
chain according to the Newman-Watts construction
\cite{newman_pla99} with some minor changes. It means that links
are added without removing any of the original ones and each site
has a maximum of five neighbors. The modified structure is
characterized by the ratio $p$ of the added and original links.
The limitation of the maximum number of neighbors prevents the
formation of large hubs and can help the (future) extension of the
generalized mean-field techniques for inhomogeneous structures.

The MC simulations are performed on systems containing $N=10^5 -
10^6$ sites. In many cases the simulations are started from a
random initial state and after a suitable transient time we have
determined the average density $\rho$ of cooperators by averaging
over a sampling time (with duration comparable to the transient
time). The transient time varied from $2 \times 10^4$ to $10^6$ MC
steps depending on the parameters (during 1 MC step each player
has a chance once on average to modify her strategy). Besides this
traditional method we used another (more efficient) technique to
determine the critical point for the transitions M $\rightarrow$ C
and D $\rightarrow$ C. In the so-called growing seed method the
simulations are started from a state where only one defector was
present in the sea of cooperators and the survival probability of
defectors is measured (by averaging over several thousand
independent runs) at different $b$ values for a fixed $K$ (for
details see Ref.~\cite{marro_99}).

\begin{figure}[ht]
\centerline{\epsfig{file=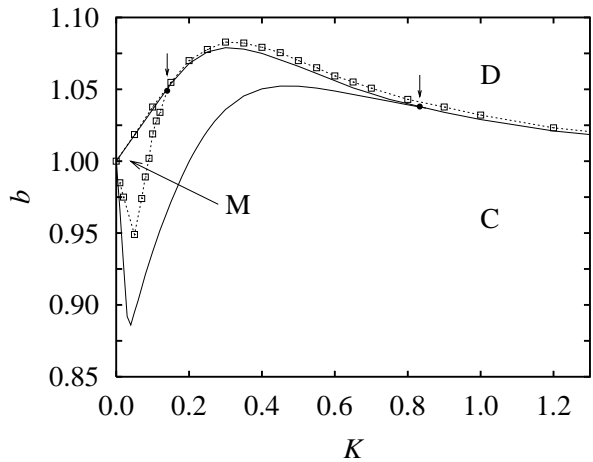,width=8cm}}
\caption{\label{fig:pd1} Phase boundaries on the $b-K$ plane for
the one-dimensional connectivity structure with $z=4$. Solid lines
indicate the prediction of the generalized mean-field technique
for $n=10$ while connected symbols stand for MC results. The
phase boundaries separate the homogeneous phases of cooperators
(C) or defectors (D), and the mixed state M where cooperators and
defectors coexist. The tricritical points are indicated by arrows
and also by full circles for the MC results and the GMF method.}
\end{figure}

Beside the MC simulations, the above model is investigated also by
the application of the generalized mean-field (GMF)
method. Within this approach the translation invariant system is
characterized by a set of configuration probabilities on a compact
cluster of $n$ neighboring sites. This approximation involves the
derivation of the hierarchy of equations of motion for the
configuration probabilities on clusters of $n$ sites (for details
see Appendix in Ref.~\cite{szabo_pr07}). Clearly, the accuracy of
this method can be improved by increasing the values of $n$. In the
present model we have found that the qualitative prediction of GMF
for the phase diagram does not change if $n > 5$. The most accurate 
approximation is achieved for $n=10$. The corresponding predictions are compared
with MC data in Fig.~\ref{fig:pd1}.

Figure \ref{fig:pd1} illustrates that both techniques show a step
like transition from C to D phase at $b=1$, that is $b_{c1}=b_{c2}=1$,
in the limit $K \to 0$. A similar behavior can be concluded if $K
\to \infty$, that is resembling the mean-field prediction
mentioned above. First order transition from C to D is found
(i.e., $b_{c1}=b_{c2} > 1$) when varying the value of temptation
$b$ for fixed noise level if its value exceeds a threshold value
($K>K^{(MC)}_{tric} \approx 0.15$). For lower noise level the
first order transition splits into two continuous critical
transitions belonging to the directed percolation universality class
\cite{marro_99} as detailed below. For this threshold value of
noise ($K=K_{tric}$) one can observe a tricritical point
($b=b^{(MC)}_{tric} \approx 1.05$) where all the mentioned three
phases can coexist.

\begin{figure}[ht]
\centerline{\epsfig{file=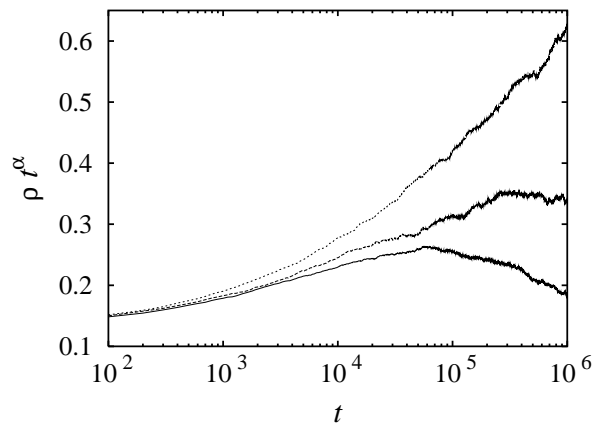,width=8cm}}
\caption{\label{fig:dp} Density decay of cooperators at $b=1.03$
for three different noise levels ($K=0.0791$, $0.0790$, and
$0.0789$ from top to bottom). The middle curve illustrates the
typical behavior in the critical point ($\alpha = 0.159$).}
\end{figure}

The continuous extinction process of both the C and D strategies
exhibits universal behavior belonging to the directed percolation
universality class \cite{janssen_zpb81}. This means that the
variation of $b$ or $K$ (as a control parameter) yields a power
law decrease of order parameter (here the density of C or D
strategies) when approaching the critical point. The algebraic
behavior is accompanied by a power law divergency in the
correlation length, relaxation time, and fluctuation of order
parameter that makes the numerical analysis time consuming
\cite{marro_99}.  In the critical point the order parameter decays
algebraically as
\begin{equation}
\label{eq:alg} \rho (t) \propto t^{-\alpha}
\end{equation}
for sufficiently long times. The value of $\alpha$ and other
critical exponents are universal (independent of the details of
dynamical rules) while their values are related to each other and
depend on the spatial dimension. For a one-dimensional stochastic
cellular automaton $\alpha = 0.159464(6)$ is found
\cite{jensen_pra91}. In order to demonstrate the same universal
behavior, the quantity $\rho t^{\alpha}$ is
plotted as a function of time in Fig.~\ref{fig:dp} where 30 independent runs of $N = 2
\times 10^6$ system size are averaged.

Notice that in Fig.~\ref{fig:pd1} the MC results are reproduced
qualitatively well by the GMF approximation. Quantitatively good
agreement is found in the function $b_{c2}(K)$ for arbitrary value
of $K$ while for the lower phase boundary the reproduction is not
satisfactory (if $K < K_{tric}$). The relevance of long-range
correlations in the extinction process of D strategies can explain
the difference in the prediction of GMF and MC results.

The $b-K$ phase diagram demonstrates another striking feature of
the evolutionary PD game for this connectivity structure. Namely,
Fig.~\ref{fig:pd1} refers to the existence of an optimum noise
level providing the best condition for the cooperators to survive.
At the same time, when increasing $K$ for $b=1.03$ one can observe
three consecutive phase transitions, namely, D $\to$ M $\to$ C
$\to$ D. The smoothed version of this behavior was reported for
those connectivity structures that cannot be spanned by
overlapping triangles supporting the spreading of cooperation in
the low noise limit for the pairwise comparison dynamical rule
\cite{szabo_pre05}.

Interestingly, one can observe another re-entrance transition
along the boundary separating the C and M phases in
Fig.~\ref{fig:pd1} that is similar to previously observed
coherence resonance phenomenon for the PD game \cite{perc_njp06a}
or even for other games \cite{traulsen_prl04}.

As mentioned above the cooperation can be maintained even at the
low noise limit (if $b > 1$) for many other connectivity
structures spanned by overlapping triangles. The present study
has demonstrated that this general trend is defeated by the
additional one-dimensional constraint providing that propagating
fronts cannot avoid each other on this structure. However, the latter
feature is destroyed when new links are added to the connectivity
structure as demonstrated in Fig.~\ref{fig:struct}.

\begin{figure}[t]
\centerline{\epsfig{file=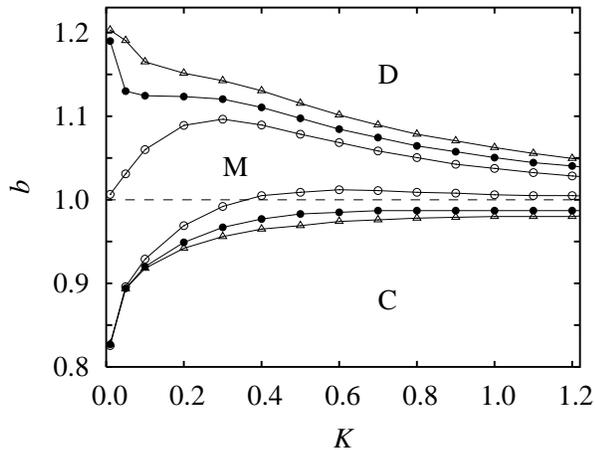,width=8cm}}
\caption{\label{fig:pd2} The lower $b_{c1}(K)$ and upper
$b_{c2}(K)$ critical points on the Newman-Watts structure for
different $p$ values. The symbols correspond to $p = 0.02$ (open
circle), $0.06$ (full circle), and $0.10$ (triangle). Solid lines
are guide to the eye and dashed line indicates a base line at
$b=1$.}
\end{figure}

The effect of new links on the $b-K$ phase diagram is illustrated
in Fig.~\ref{fig:pd2} for different values of $p$ characterizing
the portion of added links. From the present numerical results one
can conclude that the region of the M state becomes wider as $p$
increases and persists in the whole range of $K$ we studied.
Simultaneously the tricritical point disappears immediately and
the lower phase boundary $b_{c1}(K)$ remains less than 1 if $K \to
0$ as we introduce the additional links. The gradual variations
are related to the emergence of the small-world character and
simultaneously to the destruction of the strict one-dimensional
properties when $p$ is increased. On the contrary, the upper phase
boundary (between the D and M phases) tends to $b=1$ as $K$ goes
to $0$ if $p$ does not exceed a threshold value $p_{tr}= 0.04(2)$.
Unfortunately, the more accurate determination of $p_{tr}$ is
prevented by the long relaxation time diverging if $K \to 0$.

To sum up, we have systematically studied the effect of noise $K$
and temptation $b$ (to choose defection) on the measure of
cooperation for an evolutionary Prisoner's Dilemma game on a
one-dimensional chain with nearest and next-nearest neighbor
interactions.
The application of this connectivity structure is motivated by the claim 
to clarify the possible impacts of elementary topological properties.
Here the conflicting topological features result in a curious $b-K$
phase diagram where three phases can exist at special values of parameters.
It is also demonstrated that the basic features of the phase diagram
are modified drastically when new links are added to
the connectivity structure according to the Newman-Watts
construction.

\begin{acknowledgments}
This work was supported by the Hungarian National Research Fund
(Grant No. T-47003) and by the European Science Foundation (COST
P10).
\end{acknowledgments}

\end{document}